\documentclass{article}

\usepackage[preprint, nonatbib]{neurips_2021}

\usepackage[utf8]{inputenc} % allow utf-8 input
\usepackage[T1]{fontenc}    % use 8-bit T1 fonts
\usepackage{hyperref}       % hyperlinks
\usepackage{url}            % simple URL typesetting
\usepackage{booktabs}       % professional-quality tables
\usepackage{amsfonts}       % blackboard math symbols
\usepackage{nicefrac}       % compact symbols for 1/2, etc.
\usepackage{microtype}      % microtypography
\usepackage{xcolor}         % colors

\usepackage{geometry}
\usepackage{setspace}
\usepackage{cite}
\usepackage{wrapfig}
	
\usepackage{fancyhdr}
\usepackage{amssymb}
\usepackage{amsmath}
\usepackage{latexsym}
\usepackage{amsmath}
\usepackage{amscd}
\usepackage{mathtools}
\usepackage{mathrsfs}
\usepackage{graphicx}
\usepackage{enumerate}
\usepackage{dsfont}
\usepackage{tikz}
\usepackage{extarrows}
\usepackage[makeroom]{cancel}
\usepackage{blkarray}
\usepackage{hyperref}
\usepackage{multirow}
\usepackage{multicol}
\usepackage[vcentermath,enableskew]{youngtab}
\usepackage{young}

\usepackage{listings}
\usepackage{color}

\definecolor{dkgreen}{rgb}{0,0.6,0}
\definecolor{gray}{rgb}{0.5,0.5,0.5}
\definecolor{mauve}{rgb}{0.58,0,0.82}

\newcommand{\nc}{\newcommand}

\nc{\DMO}{\DeclareMathOperator}
\DMO{\id}{id}
\DMO{\rk}{rank}
\DMO{\sgn}{sgn}
\DMO{\alt}{Alt}
\DMO{\len}{len}
\DMO{\im}{img}
\DMO{\tr}{tr}

\nc{\comment}[1]{}
\nc{\trm}{\textrm}

\nc{\p}[2]{\frac{\partial #1}{\partial #2}}
\nc{\der}[2]{\frac{d #1}{d #2}}
\nc{\dd}[2]{\frac{d^2 #1}{d #2 ^2}}
\nc{\pp}[2]{\frac{\partial^2 {#1}} {\partial {#2} ^2}}
\nc{\pmix}[3]{\frac{\partial^2 {#1}}{\partial {#2} \, \partial {#3}}}
\nc{\rd}{\trm{d}}
\nc{\pa}{\partial}
\nc{\va}[2]{\frac{\del #1}{\del #2}}

\nc{\f}{\frac}

\nc{\N}{\mathds{N}}
\nc{\Z}{\mathds{Z}}
\nc{\Q}{\mathds{Q}}
\nc{\R}{\mathds{R}}
\nc{\C}{\mathds{C}}

\nc{\prob}{\mathds{P}}
\nc{\E}{\mathds{E}}
\nc{\var}{\trm{Var}}
\nc{\cov}{\trm{Cov}}
\nc{\cor}{\trm{Cor}}
\nc{\lh}{\mathcal{L}}

\nc{\St}{\mathds{S}}

\nc{\too}{\longrightarrow}
\nc{\rez}{\Rightarrow}
\nc{\bs}{\ensuremath{\blacksquare}}
\nc{\done}{\hfill\bs}
\nc{\bsh}{\backslash}
\nc{\w}{\wedge}
\nc{\til}{\tilde}
\nc{\xra}{\xrightarrow}
\nc{\xrez}{\xRightarrow}
\nc{\xle}{\xlongequal}

\nc{\union}{\bigcup}
\nc{\intt}{\bigcap}

\nc{\x}{\times}
\nc{\ox}{\otimes}
\nc{\op}{\oplus}

\nc{\al}{\alpha}
\nc{\be}{\beta}
\nc{\eps}{\varepsilon}
\nc{\epsi}{\epsilon}
\nc{\del}{\delta}
\nc{\D}{\Delta}
\nc{\Del}{\nabla}
\nc{\ph}{\varphi}
\nc{\Lam}{\Lambda}
\nc{\lam}{\lambda}
\nc{\G}{\Gamma}
\nc{\g}{\gamma}
\nc{\sig}{\sigma}
\nc{\tta}{\theta}
\nc{\Tta}{\Theta}
\nc{\kap}{\kappa}
\nc{\om}{\omega}

\nc{\lie}{\pounds}

\nc{\contr}{%
\begin{tikzpicture}[rotate=45,x=0.5ex,y=0.5ex]
\draw[line width=.2ex] (0,2) -- (3,2) (0,1) -- (3,1) (1,3) -- (1,0) (2,3) -- (2,0);
\end{tikzpicture}
}

\usetikzlibrary{automata,chains}

\nc{\Litems}[1]{\begin{enumerate}[{\bf a)}] #1 \end{enumerate}}
\nc{\Nitems}[1]{\begin{enumerate}[{\bf 1)}] #1 \end{enumerate}}
\nc{\Ritems}[1]{\begin{enumerate}[{\bf i)}] #1 \end{enumerate}}
\nc{\Items}[1]{\begin{itemize} #1 \end{itemize}}
\nc{\cntr}[1]{\begin{center} #1 \end{center}}

\nc{\bx}{\mathbf{x}}
\nc{\bz}{\mathbf{z}}
\nc{\bl}{\mathbf{l}}
\nc{\bW}{\mathbf{w}}
\nc{\bxi}{\boldsymbol\Xi}
\nc{\bI}{\mathbf{i}}
\nc{\bJ}{\mathbf{J^s}}
\nc{\bj}{\mathbf{J}}
\nc{\bM}{\mathbf{M}}
\nc{\loss}{\mathcal{L}}

\nc{\comm}[1]{{\color{red} #1}}

\nc{\floor}[1]{\left\lfloor {#1} \right\rfloor}
\nc{\ceil}[1]{\left\lceil {#1} \right\rceil}

\title{Implementing Permutations in the Brain \\ and SVO Frequencies of Languages}

\author{%
  Denis Turcu \\
  Neurobiology and Behavior\\
  Columbia University\\
  New York, NY 10027\\
  \texttt{d.turcu@columbia.edu} \\
   \And
   Christos H. Papadimitriou \\
   Department of Computer Science \\
   Columbia University \\
   New York, NY 10027 \\
   \texttt{christos@columbia.edu} \\
}

\begin{document}

\maketitle

\begin{abstract}
  The subject-verb-object (SVO) word order prevalent in English is shared by about $42\%$ of world languages. Another $45\%$ of all languages follow the SOV order, $9\%$ the VSO order, and fewer languages use the three remaining permutations. None of the many extant explanations of this phenomenon take into account the difficulty of implementing these permutations in the brain. We propose a plausible model of sentence generation inspired by the recently proposed Assembly Calculus framework of brain function. Our model results in a natural explanation of the uneven frequencies. Estimating the parameters of this model yields predictions of the relative difficulty of dis-inhibiting one brain area from another. Our model is based on the standard syntax tree, a simple binary tree with three leaves. Each leaf corresponds to one of the three parts of a basic sentence. The leaves can be activated through lock and unlock operations and the sequence of activation of the leaves implements a specific word order. More generally, we also formulate and algorithmically solve the problems of implementing a permutation of the leaves of any binary tree, and of selecting the permutation that is easiest to implement on a given binary tree.
\end{abstract}

%%%%%%%%%%%%%%%%%%%%%%%%%%%%%%%%%%%%%%%%%%%%%%%%%%%%%%%%%%%%%%%%%%%%%%%%%%%%%%%%%%%%%%%%%%%%%%%%
%%%%%%%%%%%%%%%%%%%%%%%%%%%%%%%%%%%%%%%%%%%%%%%%%%%%%%%%%%%%%%%%%%%%%%%%%%%%%%%%%%%%%%%%%%%%%%%%
%%%%%%%%%%%%%%%%%%%%%%%%%%%%%%%%%%%%%%%%%%%%%%%%%%%%%%%%%%%%%%%%%%%%%%%%%%%%%%%%%%%%%%%%%%%%%%%%
%%%%%%%%%%%%%%%%%%%%%%%%%%%%%%%%%%%%%%%%%%%%%%%%%%%%%%%%%%%%%%%%%%%%%%%%%%%%%%%%%%%%%%%%%%%%%%%%
%%%%%%%%%%%%%%%%%%%%%%%%%%%%%%%%%%%%%%%%%%%%%%%%%%%%%%%%%%%%%%%%%%%%%%%%%%%%%%%%%%%%%%%%%%%%%%%%

\section{Introduction}

In English, the subject of a sentence generally comes before the verb while the object, if present, follows both: ``dogs chase cats''. This ordering is not universal, as many languages adopt any of the six possible orderings \cite{Dryer2005, Maurits2010, Maurits2011, Maurits2014, Gell-Mann2011, Krupa1982, Hammarstrom2015, Huber2013, Aitchison1986, Tomlin1988, Greenberg1963}. The same order as in English, denoted SVO, is prevalent in French, Hebrew, modern Greek and Romanian, and overall in about $42\%$ of world languages.  The order SOV is slightly more common, accounting for $45\%$ of languages, including Hindi, Urdu, Japanese, Latin, and ancient Greek.  The orders VSO ($9\%$), VOS ($2\%$) and OVS ($1\%$) are much less common, while the order OSV ($<1\%$) is practically disregarded. In English, changing the language's SVO order creates either meaningless sentences (``chase cats dogs'') or changes the meaning (``cats chase dogs'').  In other languages, such as German, Russian, or modern Greek, deviations from the standard order are tolerated, because nouns have a {\em case} in these languages, which makes their syntactic role (subject {\em vs} object) easy to identify independently of position.  However, all languages seem to have a dominant, default word order.

There is extensive literature on justifying the widely varying frequencies of basic word orders. These past explanations are based on plausible linguistic principles related to the ease of communicating meaning, or the difficulty of learning grammar, while more recent explanations consider the mutability and evolution of word orders in languages \cite{Gell-Mann2011, Krupa1982, Tomlin1988, Greenberg1963, Maurits2010, Maurits2011, Maurits2014}.  Here we  propose that differences in the {\em difficulty of generating sentences in the brain} may play a major role in determining the relative prevalence of basic word orders.  

Language emerges in the human brain through the activity of neurons, synapses, and neural circuits. Properties of language, such as the basic word orders, are undoubtedly affected and constrained by the structure and function of the language-processing brain areas. In the past decade, there has been tremendous expansion of our experimental insight into the ways whereby the human brain processes and generates language (see \cite{Friederici2017} for a book-length exposition).  However, we are not aware of hypotheses about the function of the language organ providing insights into the origins of linguistic phenomena, such as the basic word orders.

\begin{figure}[t]
    \centering \includegraphics[scale=0.3]{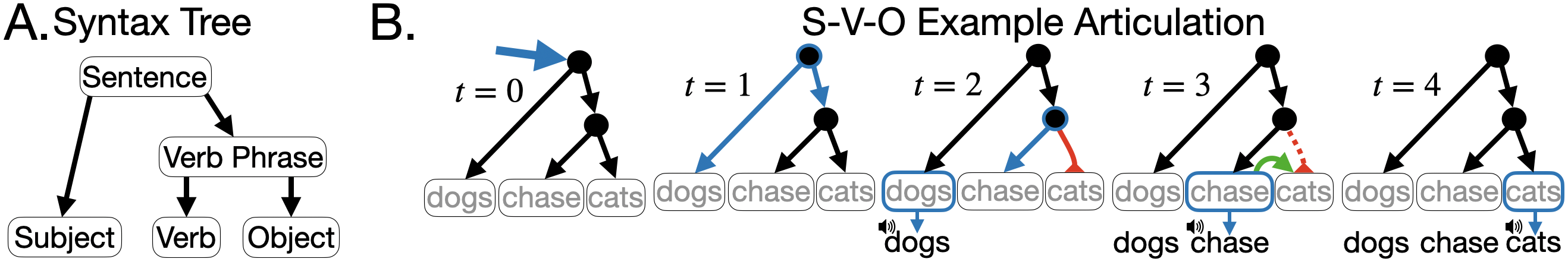}
    \vskip 0.5cm
    \centering \includegraphics[scale=0.38]{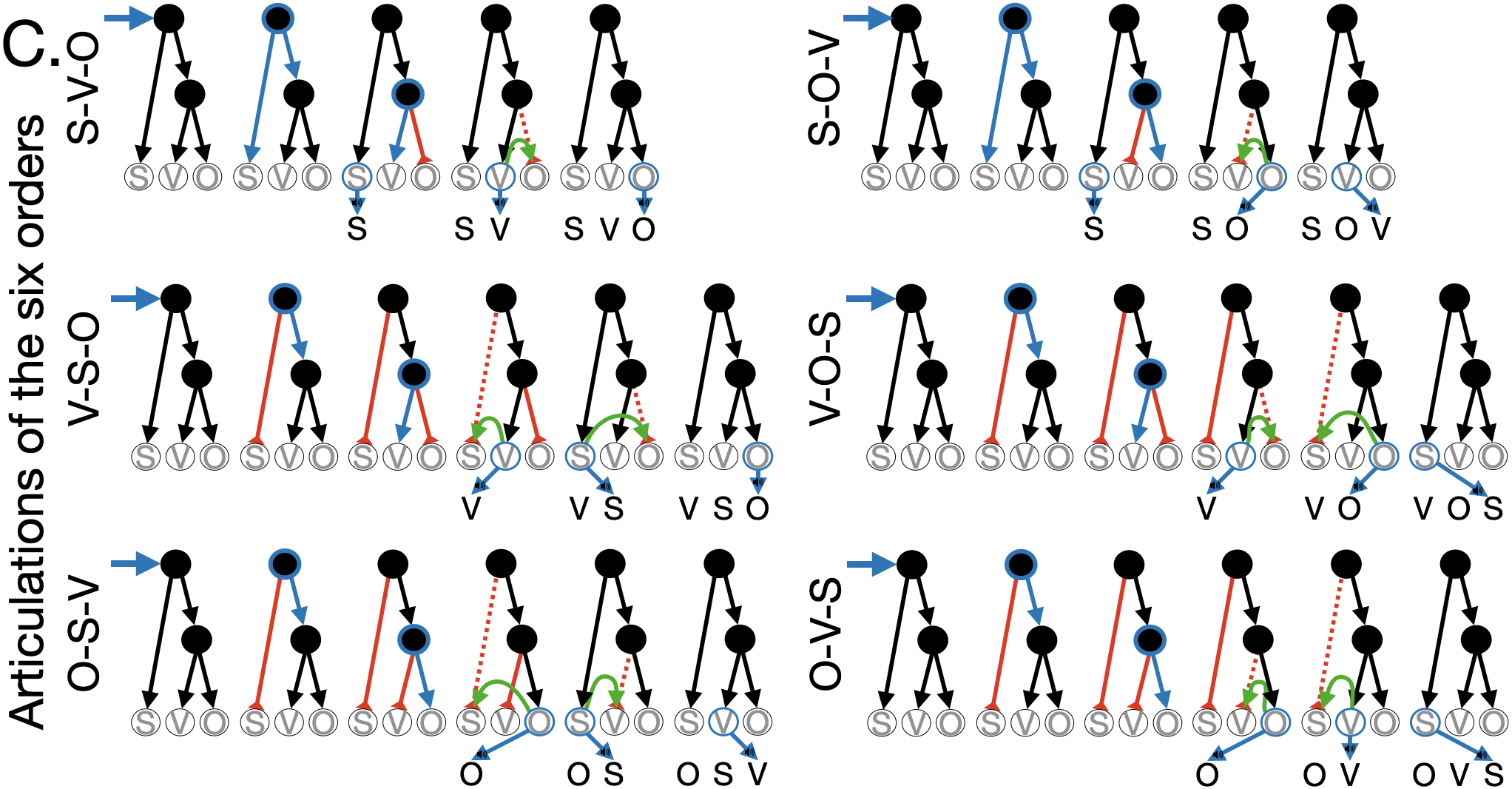}
    \caption{\small 
    {\bf (A)} Basic syntactic tree with a ``Verb Phrase" internal node and ``Subject", ``Verb" and ``Object" leaves.
    {\bf (B)} Example articulation from the syntactic tree to sequential speech for the SVO order. Black arrows are inactive. Blue arrows activate the object they point to on the next time step. Red inhibitory signals maintain a lock on the object they point to. Green arrows remove the lock.
    {\bf (C)} Articulations of all possible basic word orders to sequential speech, starting from the same syntactic tree. Appropriate lock and unlock operations dictate the basic word order.
    }
    \label{fig:fig1}
\end{figure}

Exploring the neural basis of cognitive functions, such as language, is daunting because of the current large gap in scale between high-level models of the brain used in cognitive science and the low-level models of neurons, synapses and circuits common in neuroscience. A computational system of intermediate scale is necessary for bridging this gap. Such a system was recently introduced, the {\em Assembly Calculus (AC) } \cite{Papadimitriou2020}. This system's fundamental computational object is the {\em assembly of neurons} -- a large set of excitatory neurons representing a real world object or concept, such as a word. The AC creates and manipulates assemblies of neurons using a set of operations, e.g. project, associate, merge, etc. These operations can be shown to be realizable by the activity of real neurons and synapses through mathematical proofs, simulations in an idealized model, as well as in models of realistic spiking neurons. They can also arguably implement cognitive functions, such as language generation and comprehension. In particular, \cite{Papadimitriou2020} proposes a program in the AC that is capable of generating simple sentences, such as ``dogs chase cats'', by building a basic syntactic tree of assemblies in Broca's area, whose leaves are representations of the individual words in Wernicke's area. It is noted in \cite{Papadimitriou2020} that, once such a tree has been generated, the person may choose to {\em articulate} the associated sentence, that is, to turn the tree into sequential speech [Figure \ref{fig:fig1} A \& B]. At this stage, word order becomes relevant -- whereas generation of the tree is order-independent.

Here we propose a simple computational model for sequentially articulating the words of a sentence, starting from the basic syntactic tree. We show that the words, which are stored in the leaves of the sentence-tree [Figure \ref{fig:fig1} A], can be ordered in the brain through Assembly Calculus operations inhibiting and dis-inhibiting specific brain areas (called {\em lock and unlock operations} in this paper). We point out that this simple model immediately predicts the prevalence of the two orders SOV and SVO over the other four, a direct consequence of the asymmetries inherent in the standard syntax tree [Figure \ref{fig:fig1} C], whose root represents the Sentence, internal node represents the Verb Phrase and leaves represent Subject, Verb and Object. Linguistic arguments \cite{Maurits2011, Dunn2011, Tomlin1988, Krupa1982} as well as experimental evidence \cite{frankland2020} are in favor of this particular form of the basic syntax tree.
In addition, we show that a refinement of our model can recover, through appropriate parameters, the precise frequencies of the six orders.

%%%%%%%%%%%%%%%%%%%%%%%%%%%%%%%%%%%%%%%%%%%%%%%%%%%%%%%%%%%%%%%%%%%%%%%%%%%%%%%%%%%%%%%%%%%%%%%%
%%%%%%%%%%%%%%%%%%%%%%%%%%%%%%%%%%%%%%%%%%%%%%%%%%%%%%%%%%%%%%%%%%%%%%%%%%%%%%%%%%%%%%%%%%%%%%%%
%%%%%%%%%%%%%%%%%%%%%%%%%%%%%%%%%%%%%%%%%%%%%%%%%%%%%%%%%%%%%%%%%%%%%%%%%%%%%%%%%%%%%%%%%%%%%%%%
%%%%%%%%%%%%%%%%%%%%%%%%%%%%%%%%%%%%%%%%%%%%%%%%%%%%%%%%%%%%%%%%%%%%%%%%%%%%%%%%%%%%%%%%%%%%%%%%
%%%%%%%%%%%%%%%%%%%%%%%%%%%%%%%%%%%%%%%%%%%%%%%%%%%%%%%%%%%%%%%%%%%%%%%%%%%%%%%%%%%%%%%%%%%%%%%%

\section{The Model}
Our computational model is based on the basic operations of the AC, summarized here (see \cite{Papadimitriou2020} for mode details). An extant assembly $x$ in a brain area $\mathcal{A}$ can be {\em projected} in another brain area $\mathcal{B}$ to create a new assembly $y$. The new assembly is endowed with strong synaptic connectivity from $x$ and hence will theretofore fire whenever $x$ fires. Two assemblies $x$ and $y$ from two different brain areas, $\mathcal{A}$ and $\mathcal{B}$, can create a {\em merge} assembly $z$ in a third brain area, $\mathcal{C}$, with strong synaptic connectivity {\em back and forth} from $x$ and $y$. The merge assembly, $z$, can be henceforth considered to be the parent of $x$ and $y$ in a tree thus constructed. This process can be repeated to build arbitrary binary trees -- for example, the basic syntax tree of a sentence.

A short sequence of AC operations can {\em generate sentences.} A simple sentence such as ``dogs chase cats'' can be generated by first identifying the three assemblies corresponding to the three words in the lexicon, believed to reside in the medial temporal lobe \cite{Friederici2017}. Then, these word-assemblies project to create three new assemblies within separate subareas of Wernicke's area in the Superior Temporal Gyrus, corresponding to Subject, Verb and Object brain areas. Next, the Verb and Object assemblies (in this example corresponding to ``chase'' and ``cats'', respectively) merge to create a {\em Verb Phrase} assembly in Broca's area \cite{Friederici2017}. Finally, the Subject and Verb Phrase assemblies merge to create an assembly representing the Sentence [Figure \ref{fig:fig1}], in another subarea of Broca's area \cite{Friederici2017}.  A sentence may have many other constituents, such as determiners, adjectives, adverbs, and propositional phrases, but we focus only on the tree built from its three basic syntactic parts: Subject, Verb, and Object (see Section \ref{sec:tree_problem} for a treatment of the general problem).

Three different binary trees can be built from three leaves, by grouping any two of these leaves first.  There is a broad consensus in Linguistics \cite{Maurits2011, Dunn2011, Tomlin1988, Krupa1982}, as well as evidence from cognitive experiments \cite{frankland2020}, supporting the basic tree described above [Figure \ref{fig:fig1} A] with an internal Verb Phrase node whose constituents are Verb and Object.

Once the sentence is generated, it may be {\em articulated}, that is, converted into speech. This can be done by exciting the root of the tree -- the Sentence assembly -- which then will excite its children in the tree and so on.  Eventually, all three leaves will be excited. Each leaf can mobilize motor programs which will articulate each word, but this must be done sequentially. Therefore, {\em one of the six orders must be selected and implemented.}

Perhaps the simplest and most biologically realistic mechanism for implementing a particular order involves two plausible primitives, which we call {\em lock} and {\em unlock}. These primitives correspond to the familiar neural processes of inhibition of an area (the activation of a population of inhibitory neurons which will prevent excitatory neurons in this area from firing) and dis-inhibition (the inhibition of the inhibitory population) \cite{Buzsaki2007, Jackson2015, Letzkus2015}. In particular, upon firing, an assembly in the tree can inhibit one of its children from firing, assuming that the child is a leaf. Secondly, any leaf can, upon firing, dis-inhibit any other leaf.

Sentence articulation arises from a syntax tree whose (internal) nodes and leaves are equipped with appropriate lock and unlock operations. Generally, nodes activate their children on the next time-step upon firing; however, nodes can lock any of its {\em leaf children} for any number of time-steps. A leaf can unlock a single other leaf which is being locked by a node. This process makes the inhibited leaf fire at the next time-step. A particular realizable choice of lock and unlock operations determines the basic word order articulated from the basic syntax tree; Figure \ref{fig:fig1} C shows the choices with the smallest possible number of lock and unlock operations implementing the six orders.

\paragraph{Basic Model} Our basic model is based on the observation that among the six orders, only two can be implemented by just one lock and one unlock operation, whereas all others require two lock and two unlock operations [Figure \ref{fig:fig1} C]. In particular, this model deliberately ignores the relative difficulty of various lock operations, or the additional difficulty of maintaining an operation for multiple time-steps by sustained inhibition over long times.

\paragraph{Extended Model} The extended version incorporates detailed metabolic costs of implementing each order. Metabolic costs of brain function have been well documented \cite{Buzsaki2007, Laughlin1998, Laughlin2001, Attwell2001, Niven2008}, in particular in relation to differences between excitatory and inhibitory energy consumption \cite{Buzsaki2007, Attwell2001} and in relation to information transmission \cite{Laughlin1998, Laughlin2001} and sustained neural activity \cite{Buzsaki2007}. Our extended model integrates this evidence using parameters which stand for specific costs associated with each operation and the number of time-steps which maintained that operation. That is, we propose a parametrized model, whose parameters represent energy-time costs of certain brain functions.  We note here that, instead of, or in addition to, metabolic costs, one could consider the costs involved in {\em learning} to implement the sequence of brain operations required; however, since such costs are difficult to assess without further assumptions we do not pursue this approach here, except to note that the basic model arguably models it adequately: a sequence of two operations must be easier to learn than a sequence of four.

%%%%%%%%%%%%%%%%%%%%%%%%%%%%%%%%%%%%%%%%%%%%%%%%%%%%%%%%%%%%%%%%%%%%%%%%%%%%%%%%%%%%%%%%%%%%%%%%
%%%%%%%%%%%%%%%%%%%%%%%%%%%%%%%%%%%%%%%%%%%%%%%%%%%%%%%%%%%%%%%%%%%%%%%%%%%%%%%%%%%%%%%%%%%%%%%%
%%%%%%%%%%%%%%%%%%%%%%%%%%%%%%%%%%%%%%%%%%%%%%%%%%%%%%%%%%%%%%%%%%%%%%%%%%%%%%%%%%%%%%%%%%%%%%%%
%%%%%%%%%%%%%%%%%%%%%%%%%%%%%%%%%%%%%%%%%%%%%%%%%%%%%%%%%%%%%%%%%%%%%%%%%%%%%%%%%%%%%%%%%%%%%%%%
%%%%%%%%%%%%%%%%%%%%%%%%%%%%%%%%%%%%%%%%%%%%%%%%%%%%%%%%%%%%%%%%%%%%%%%%%%%%%%%%%%%%%%%%%%%%%%%%

\section{Complexity of Generation Explains SVO Frequencies}

\paragraph{Basic Model} The basic model immediately predicts the prevalence of the SVO and SOV basic word orders. Each choice of articulation order requires a number of inhibition and dis-inhibition operations. Importantly, inhibition and dis-inhibition primitives are known to require significant brain energy consumption \cite{Buzsaki2007, Laughlin1998, Attwell2001}. Furthermore, each additional operation makes the articulation program longer and more complex, and hence presumably renders this aspect of language more difficult for the learner.  The orders SVO or SOV can be generated with one lock and one unlock operation, while the other four orders need two lock and two unlock operations [Figure \ref{fig:fig1} C]. Therefore, the prevalence of the SVO and SOV basic word orders arises spontaneously from this account.

\begin{figure}[t]
    \centering
    \centering \includegraphics[scale=0.31]{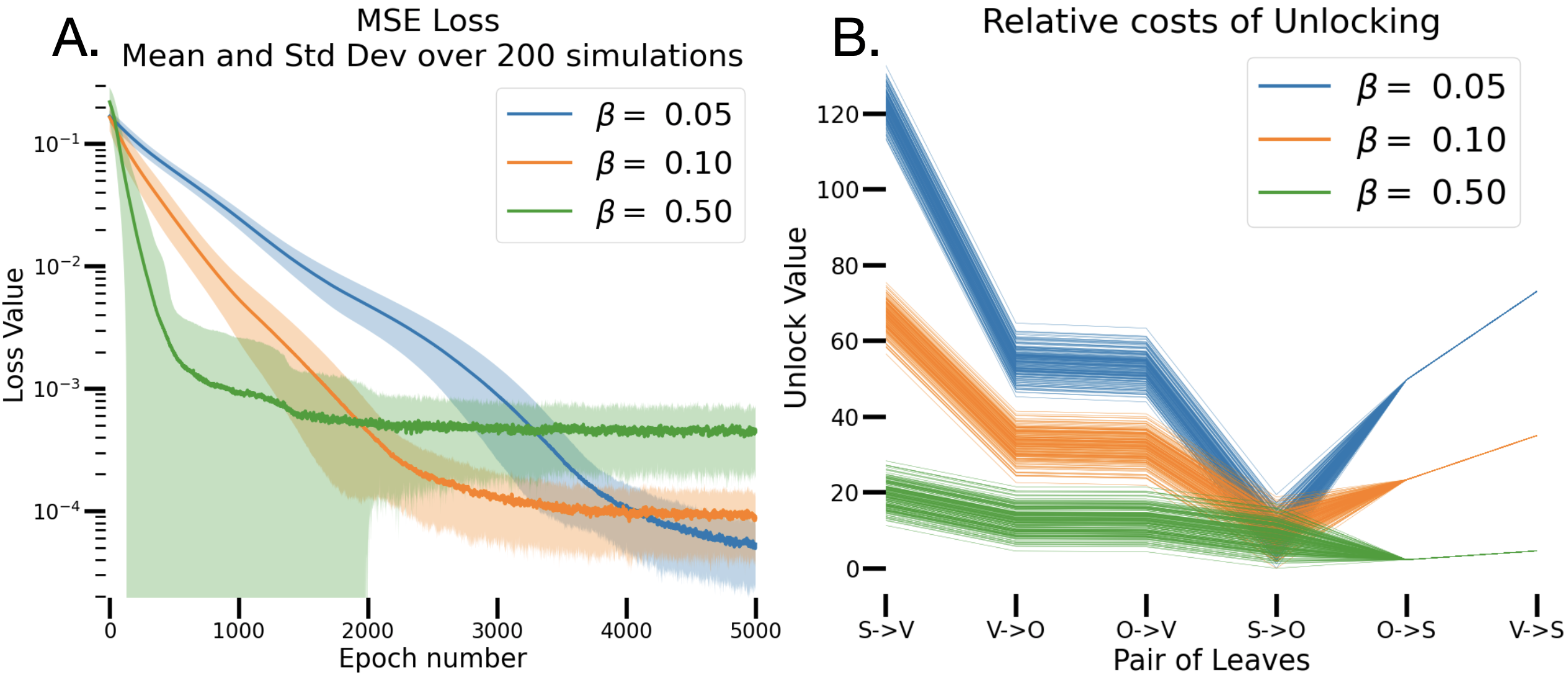}
    \caption{\small 
    {\bf (A)} Loss value plotted during the epochs of gradient descent for various $\be$ values. The lines represent the average loss and the shaded areas the standard deviation over $200$ initializations.
    {\bf (B)} Relative costs of the unlock operation from one leaf to another. Colors represent models with varying $\be$. Each line represents the optimized parameters for one model. Note the degeneracy of the solutions for the first four leaf pairs: the lines differ only by an additive constant.
    }
    \label{fig:fig2}
\end{figure}

A na\"ive statistical-mechanical argument applied to our basic model qualitatively predicts the frequencies of the basic word orders. This argument [Section \ref{sec:stat_mech}, \ref{sec:stat_mech_basic}] is based on the Boltzmann distribution \cite{sethna2006}, in which states with energy level $L$ are prevalent with probability proportional to $e^{-\beta L}$, for a constant $\beta$.  For simplicity, we take $\beta=1$ in this account (but in our experiments for the extended model we use a wide range of values for $\beta$ [Figure \ref{fig:fig2}]). The states of our model are the six basic word orders and the associated energies are the number of operations required by each articulation choice. The optimal choices for SVO or SOV have low energies, requiring only {\em two} operations (one lock and one unlock), while the other four optimal choices have high energies, requiring {\em four} operations (two lock and two unlock) [Figure \ref{fig:fig1} C]. The prevalence of the six orders SVO, SOV, VSO, VOS, OSV, and OVS would be proportional to the numbers $e^{-2}, e^{-2}, e^{-4}, e^{-4}, e^{-4}, e^{-4}$, respectively. The orders SVO and SOV would then be expected to be more frequent than the rest by a factor of $e^2\approx 7.4$, which qualitatively matches the observations. Our basic model predicts frequencies of approximately $39\%$ on average for SVO and SOV, and $5.5\%$ on average for the rest, while the true average frequencies are $43.5\%$ and, respectively $3.25\%$).

\paragraph{Extended Model} The extended model can recover the precise frequencies of the six basic word orders by introducing more parameters and fitting them to the observed data; it is robust to various hyper-parameters, including $\beta$. The extended model also takes into account the cost of the sustained lock operations over a period of time-steps, whose cost is a variable hyper-parameter [Section \ref{sec:stat_mech_extended}]. The remaining six parameters represent the different costs of the unlock operations between each of the six pairs of leaves of the syntax tree. These six costs form the basis of a system of six non-linear equations; these equations arise from matching the Boltzmann distribution model for the six basic word orders with the respective observed frequencies [each term in Equation \ref{eq:loss}]. We note that the equations display an analytical degeneracy which is also recovered from  the simulations; specifically, four of the six parameters can only be found up to a common additive constant [Section \ref{sec:stat_mech_extended}]. This degeneracy is manifest in Equation \ref{eq:sorted_Us}, in that these four parameters cannot be compared with the other two.

The system of equations does not have an analytical solution, but the six parameters can be approximated using gradient optimization. This method finds the same qualitative results for different values of the coefficient $\beta$ [Figure \ref{fig:fig2} A] and for different values of other hyper-parameters. The results of these calculations are robust enough to support certain predictions about the relative costs of dis-inhibiting one brain area from another [Figure \ref{fig:fig2} B]. More specifically, we find that:
\begin{equation}
    \label{eq:sorted_Us}
    U_{S\to V} > U_{V\to O} \gtrapprox U_{O\to V} > U_{S\to O} \;\;\;\trm{ and }\;\;\; U_{V\to S} > U_{O\to S},
\end{equation}
where $U_{x\to y}$ is the cost to dis-inhibit assembly $y$ from assembly $x$. 

%%%%%%%%%%%%%%%%%%%%%%%%%%%%%%%%%%%%%%%%%%%%%%%%%%%%%%%%%%%%%%%%%%%%%%%%%%%%%%%%%%%%%%%%%%%%%%%%
%%%%%%%%%%%%%%%%%%%%%%%%%%%%%%%%%%%%%%%%%%%%%%%%%%%%%%%%%%%%%%%%%%%%%%%%%%%%%%%%%%%%%%%%%%%%%%%%
%%%%%%%%%%%%%%%%%%%%%%%%%%%%%%%%%%%%%%%%%%%%%%%%%%%%%%%%%%%%%%%%%%%%%%%%%%%%%%%%%%%%%%%%%%%%%%%%
%%%%%%%%%%%%%%%%%%%%%%%%%%%%%%%%%%%%%%%%%%%%%%%%%%%%%%%%%%%%%%%%%%%%%%%%%%%%%%%%%%%%%%%%%%%%%%%%
%%%%%%%%%%%%%%%%%%%%%%%%%%%%%%%%%%%%%%%%%%%%%%%%%%%%%%%%%%%%%%%%%%%%%%%%%%%%%%%%%%%%%%%%%%%%%%%%

\section{A Statistical-Mechanical Argument} \label{sec:stat_mech}
In statistical mechanics, the probability of a given state of a system depends on its energy and ambient temperature. The Boltzmann distribution provides a way to estimate the thermal equilibrium configuration of all the states of a system. The probability of a state with energy $E_i$ is proportional to:
\begin{equation}
    p_i \propto \exp{(-\be E_i)},
\end{equation}
where $\be$ is a scale factor, inversely proportional to temperature. The system we describe has six states, therefore, the probability of each state is:
\begin{equation}
    \label{eq:probs}
    p_i = \f{\exp{(-\be E_i)}}{\sum_j\exp{(-\be E_j)}},
\end{equation}
for $i,j \in \{SVO, SOV, VSO, VOS, OVS, OSV\}$. 

These formulas are simply a heuristic way, aligned with physical principles, of modeling how complexity affects probabilities; however, we also note that in the neuroscience literature (see e.g.~\cite{Laughlin1998}) metabolic costs are thoroughly discussed with respect to thermal energy. On this account, we choose the energies of our states to be $E_i \sim 1$ and we assume $\be\lesssim1$.

\subsection{Basic Model} \label{sec:stat_mech_basic}
The basic model only takes into account the number of lock or unlock operations required to implement a specific order. It is clear by inspection of Figure \ref{fig:fig1} C that:
\begin{equation}\begin{cases}
    \hfill E_{SVO} = E_{SOV} &= 2, \\
    E_{VSO} = E_{VOS} = E_{OVS} = E_{OSV} &= 4. \\
\end{cases}\end{equation}
The probabilities of occurrence of each order are then:
\begin{equation}\begin{cases}
    \hfill p_{SVO} = p_{SOV} &= \f{\exp{(-2)}}{2\exp{(-2)} + 4\exp{(-4)}} \approx 0.39 ,\\
    p_{VSO} = p_{VOS} = p_{OVS} = p_{OSV} &= \f{\exp{(-4)}}{2\exp{(-2)} + 4\exp{(-4)}} \approx 0.055 ,\\
\end{cases}\end{equation}
which qualitatively match the observed frequencies.

\subsection{Extended Model} \label{sec:stat_mech_extended}
The probabilities of the states are invariant with respect to an additive constant to all energies possible in the system (follows from Equation \ref{eq:probs}). Therefore, we skip global metabolic costs, such as the three leaves firing and the `R' and `VP' nodes being activated, which add to all six energies. We include them as an additive constant, together with other global costs associated to speech. We define three sets of parameters which correspond to different costs in the sentence generation process. $A_x$ is the cost to directly activate leaf $x$, where $x \in \{S, V, O\}$. $L_{x\to y}$ is the cost to maintain a lock from $x$ to $y$ for one time-step, where $x\to y \in \{R\to S,VP\to V, VP\to O\}$. $U_{x\to y}$ is the cost to unlock leaf $y$ from leaf $x$, where $x,y \in \{S, V, O\}$. By inspection from Figure \ref{fig:fig1} C, the energies of the six states are:
\begin{equation}\begin{cases}
    \label{eq:extended_energies_full}
    E_{SVO} = A_S + A_V + 2 L_{VP\to O} + U_{V\to O} + \mathcal{C}, \\
    E_{SOV} = A_S + A_O + 2 L_{VP\to V} + U_{O\to V} + \mathcal{C}, \\
    E_{VSO} = A_V + 3 L_{VP\to O} + 3 L_{R\to S} + U_{V\to S} + U_{S\to O} + \mathcal{C}, \\
    E_{VOS} = A_V + 2 L_{VP\to O} + 4 L_{R\to S} + U_{V\to O} + U_{O\to S} + \mathcal{C}, \\
    E_{OVS} = A_O + 2 L_{VP\to V} + 4 L_{R\to S} + U_{O\to V} + U_{V\to S} + \mathcal{C}, \\
    E_{OSV} = A_O + 3 L_{VP\to V} + 3 L_{R\to S} + U_{O\to S} + U_{S\to V} + \mathcal{C}. \\
\end{cases}\end{equation}

We assume that the unknown parameters of this model are the $U_{x\to y}$'s and that the activation and inhibition of different nodes have similar metabolic costs, specifically $A_x = L_{x\to y} = 1$. Plugging these in Equation \ref{eq:extended_energies_full} and subtracting a global constant, the relevant energies are:
\begin{equation}\begin{cases}
    \label{eq:extended_energies_short}
    E_{SVO} = U_{V\to O}, \\
    E_{SOV} = U_{O\to V}, \\
    E_{VSO} = U_{V\to S} + U_{S\to O} + 3, \\
    E_{VOS} = U_{V\to O} + U_{O\to S} + 3, \\
    E_{OVS} = U_{O\to V} + U_{V\to S} + 3, \\
    E_{OSV} = U_{O\to S} + U_{S\to V} + 3. \\
\end{cases}\end{equation}

The six unknown parameters cannot be found analytically, but adding an arbitrary constant to four of the parameters does not change the Boltzmann distribution of the system. In particular, adding a constant metabolic cost to $U_{V\to O}, U_{O\to V}, U_{S\to V}, U_{S\to O}$ changes all six energies by the same constant, and thus leaves probabilities unaffected. This effect appears in the simulations: $U_{V\to S}, U_{O\to S}$ are found to have fixed values, while $U_{V\to O}, U_{O\to V}, U_{S\to V}, U_{S\to O}$ are found up to a common additive constant [Figure \ref{fig:fig2} B].

The six unknown parameters can be estimated using gradient methods. The mean squared error between the probabilities of occurrence, $p_i$ from Equation \ref{eq:probs}, and observed data on the frequency of the basic word orders, called $f_i$, is minimized [Equation \ref{eq:loss}]. Gradient descent on this loss function with various initial conditions provides the solutions of our simulations. Our model is robust to various parameters, in the sense that the solutions found over broad parameter values maintain the same relative magnitude of the six main parameters as in Equation \ref{eq:sorted_Us}. Indeed, the optimization is robust with respect to various assumptions for the values of $A_x$ and $L_{x\to y}$. Moreover, $\be$ can vary across many orders of magnitude, from $0.01$ to $2$, yet smaller $\be$ makes the optimization converge faster. The simulations reach small loss values and are robust to the hyper-parameters mentioned [Figure \ref{fig:fig2} A].

\begin{equation}
    \label{eq:loss}
    loss = \f16 \sum_i (p_i - f_i)^2
\end{equation}

%%%%%%%%%%%%%%%%%%%%%%%%%%%%%%%%%%%%%%%%%%%%%%%%%%%%%%%%%%%%%%%%%%%%%%%%%%%%%%%%%%%%%%%%%%%%%%%%
%%%%%%%%%%%%%%%%%%%%%%%%%%%%%%%%%%%%%%%%%%%%%%%%%%%%%%%%%%%%%%%%%%%%%%%%%%%%%%%%%%%%%%%%%%%%%%%%
%%%%%%%%%%%%%%%%%%%%%%%%%%%%%%%%%%%%%%%%%%%%%%%%%%%%%%%%%%%%%%%%%%%%%%%%%%%%%%%%%%%%%%%%%%%%%%%%
%%%%%%%%%%%%%%%%%%%%%%%%%%%%%%%%%%%%%%%%%%%%%%%%%%%%%%%%%%%%%%%%%%%%%%%%%%%%%%%%%%%%%%%%%%%%%%%%
%%%%%%%%%%%%%%%%%%%%%%%%%%%%%%%%%%%%%%%%%%%%%%%%%%%%%%%%%%%%%%%%%%%%%%%%%%%%%%%%%%%%%%%%%%%%%%%%

\section{The Leaf Order Problem} \label{sec:tree_problem}

The problem of generating the leaves of a tree in a given order through lock and unlock steps, so that the total cost is minimized, can be generalized to arbitrary binary trees --- presumably representing full-fledged sentences, as opposed to the basic syntax tree with only the three leaves: Subject, Verb, and Object. We are not aware of a prior mention of this problem in the literature.

To define the problem, suppose that we are given a binary tree with $n$ leaves, and a permutation $\sig$ of $\{1,\ldots,n\}$ which dictates the order of the leaves firing, such that leaf $i$ fires before leaf $j$ if and only if $\sig(i) < \sig(j)$. An internal node can lock any leaf child when it fires, including locking both leaf children if it has two. A leaf can unlock any other leaf upon firing. We are interested in creating a schedule for the firings of the nodes of the tree so that the permutation $\sig$ of the leaves is realized. We want to find the schedule which requires the fewest lock and unlock operations and can be completed in the fewest number of parallel time steps --- that is, the last leaf $z$ with $\sig(z)=n$ will fire as soon as possible. We call this {\em Problem 1}.  Note that it may not be possible for the leaves to fire consecutively (imagine a tree that has two leaves $1,2$ with $\sig(1)=1, \sig(2)=2$ that are grandchildren of the root, while all other leaves have much larger depth). 

A slightly more complicated problem requires us to find the permutation $\sig$ which uses the smallest possible number of lock and unlock operations, given an arbitrary tree. We call this {\em Problem 2}.  We can prove the following results:

\paragraph{Theorem:} 
\begin{enumerate}{\em 
\item
Problem 1 can be solved in $O(n\log n)$ time through a greedy algorithm. 
\item Problem 2 can be solved by an adaptation of the same greedy algorithm, if all lock and unlock steps have unit cost. 
\item However, if the unlock steps have different costs, even if the costs are restricted to be either one or two, Problem 2 is NP-hard.}
\end{enumerate}

\paragraph{Proof:} \hfill

(1) We describe the algorithm informally. It entails the sequential firing of all nodes of the tree, starting from the root; the firing propagates from a node to its children down the tree (a breadth-first search implemented by a queue of nodes). Specifically, the root fires at the first parallel time step. At step $t+1$, the internal nodes whose parents fired at step $t$ will fire. Additionally, {\em any leaf unlocked} by another leaf at time $t$ will fire at time $t+1$ (there will be at most one unlocked leaf at any time step).  Finally, if one of the internal nodes firing has any leaf children, then each child is locked {\em unless it is the next leaf to be output}.

To keep track of leaves we maintain a separate heap of {\em locked leaves} ordered by $\sig$, initially empty, and an index {\tt next}, initially {\tt 1}. If at some step we encounter a leaf child $i$ of a node being processed, there are two cases:  If $\sig(i) =$ {\tt next}, and no other leaf has been output during this step, then the leaf is output immediately and {\tt next} is increased by {\tt 1}.  Otherwise, $\sig(i) >$ {\tt next}, and $i$ joins the heap of locked leaves. At the beginning of parallel step $t$ (the round of breadth-first-search processing the nodes of the tree at depth $t-1$), we check whether the {\tt min} of the heap, call it $m$, has $\sig(m) =$ {\tt next}. If so, then we output $m$ and increment {\tt next}.  We then proceed with the breadth-first search.  The algorithm terminates when both the heap and the queue are empty.

We claim that this algorithm outputs the leaves in the $\sig$ order, and that it does so with the fewest lock and unlock operations and in the fewest parallel steps possible. We first claim that every leaf is output as early, in terms of parallel time, as possible. This follows from two things: (a) no leaf $i$ can be output earlier than time $T(i)$, where $T(i)$ satisfies the recurrence $T(i)= \max\{T(\sig^{-1}(\sig(i)-1)+1, {\rm depth}(i)\}$ if $i$ is not the first leaf and $T(i) ={\rm depth}(i)$ otherwise; and (b) the algorithm achieves this time, as can be shown by induction on $\sig(i)$.  We also claim that it implements the permutation with the fewest locks, which follows from the two facts that (c) the minimum possible number of locks is $n-1$ minus the number of {\em coincidences,} where a coincidence is an $i$ for which the two terms in the recursive definition of $T$ above are equal, and (d) such coincidences are caught and exploited by the algorithm. 

 (2) For Problem 2, we start by noticing that every leaf $i$ becomes available to be output at time ${\rm depth}(i)$.  Second, a leaf can be output without lock/unlock steps only if it is output at the precise time it becomes available.  Otherwise, if many leaves have the same depth, all but one of them can be feasibly postponed to any time in the future, and unlocked by the leaf that was output immediately before it.  Hence the following greedy algorithm achieves the minimum number of lock/unlock steps:
We define a one-to one mapping from the $n$ leaves to the time slots $\{d,d+1,\ldots,D+n\}$, where $d$ and $D$ is the minimum and maximum depth of a leaf of the tree:  First, each leaf $i$ is mapped to ${\rm depth}(i)$, which creates a map which is not one-to-one because of collisions.  We then repeatedly go through the time slots, from smaller to larger starting from $d$ and execute the following algorithm: for any time slot $t$, if it has $\ell > 1$ leaves mapped to it, select $\ell-1$ of these leaves and assign them to the $\ell -1$ empty time slots greater than $t$ and closest to $t$, resolving ties arbitrarily.  It is easy to see that this algorithm chooses the permutation of the leaves which has the maximum number of coincidences (leaves fire exactly when they become available), in the sense of the previous paragraph, and thus the minimum possible number of lock and unlock steps.

 (3) Finally, for NP-hardness:  Imagine that the tree is a full binary tree of depth $d$ --- that is, $n=2^d$ and all leaves arrive simultaneously.  Then all permutations are available, and we need to chose the ones that order $\sig(1),\sig(2),\ldots,\sig(n)$ such that $\sum_{i=2}^n {\rm unlockcost}(\sig(i-1),\sig(i))$ is as small as possible.  It is easy to see that this is a generic instance of the (open-loop) traveling salesman problem, which is known to be NP-hard even if the lengths of the edges are either one or two \cite{Papadimitriou1993}.  This completes the proof of Part (3) and of the theorem.

%%%%%%%%%%%%%%%%%%%%%%%%%%%%%%%%%%%%%%%%%%%%%%%%%%%%%%%%%%%%%%%%%%%%%%%%%%%%%%%%%%%%%%%%%%%%%%%%
%%%%%%%%%%%%%%%%%%%%%%%%%%%%%%%%%%%%%%%%%%%%%%%%%%%%%%%%%%%%%%%%%%%%%%%%%%%%%%%%%%%%%%%%%%%%%%%%
%%%%%%%%%%%%%%%%%%%%%%%%%%%%%%%%%%%%%%%%%%%%%%%%%%%%%%%%%%%%%%%%%%%%%%%%%%%%%%%%%%%%%%%%%%%%%%%%
%%%%%%%%%%%%%%%%%%%%%%%%%%%%%%%%%%%%%%%%%%%%%%%%%%%%%%%%%%%%%%%%%%%%%%%%%%%%%%%%%%%%%%%%%%%%%%%%
%%%%%%%%%%%%%%%%%%%%%%%%%%%%%%%%%%%%%%%%%%%%%%%%%%%%%%%%%%%%%%%%%%%%%%%%%%%%%%%%%%%%%%%%%%%%%%%%

\section{Discussion}

Linguistic phenomena should be constantly reinterpreted under the light of new findings, including advancements in our understanding of language processing in the brain. Despite recent progress in this front, articulating the constraints imposed by the neural processes involved in the language function is not easy, due to a large gap, in both scale and focus, between cognitive and systems neuroscience. This work attempts to bridge this gap using an intermediary computational framework, the Assembly Calculus,  providing a new explanation of the difference in frequencies of the six basic word orders in languages in terms of the difficulty of generating an order from the basic syntax tree of the sentence.

The simplest version of our model qualitatively matches the observed basic word order frequencies, and the the most complex version can be tuned to predict the exact frequencies. However, we suspect that the latter calculation may constitute overfitting, as other considerations are likely to enter in the determination of these frequencies, including linguistic considerations of communication efficiency and learnability. These other factors were heretofore the only ones used for this purpose. Our model is not meant to replace these arguments, but add to them and it provides an additional basis for breaking the symmetry of the basic word orders. 

\begin{figure}[b]
    \centering
    \centering \includegraphics[scale=0.31]{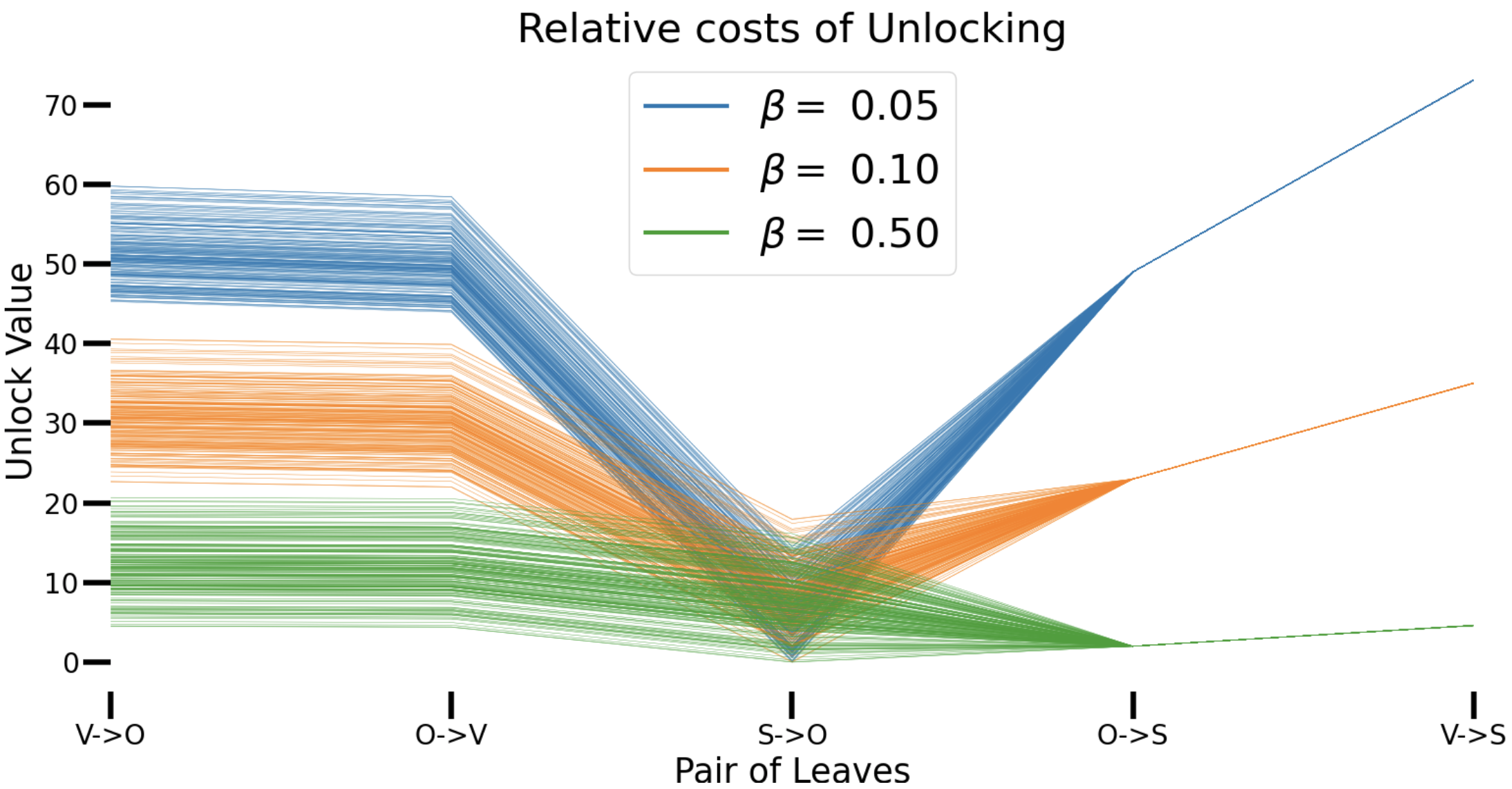}
    \caption{\small 
    The predictions if the primacy of Subject and Verb is the cause of the low frequencies of the OSV and OVS orders. Notice that there is no prediction for $U_{S\to V}$.}
    \label{fig:fig3}
\end{figure}

We believe that the ultimate explanation of the phenomenon of word orders will integrate both linguistic and neurocomputational evidence, and perhaps learnability considerations or more kinds to come. For an example of how this can be done, let us take the linguistic argument that the primary cause of the extreme rarity of orders starting with ``O'' may not be the difficulty of unlocking Subject or Verb subareas from the Object area according to our model, but the relative subsidiary semantic role of Object in a sentence, compared to the primacy of the Subject and the Verb \cite{Maurits2011, Tomlin1988, Krupa1982}.  In the face of this, we may decide that the low frequencies of the OSV and OVS orders are adequately explained on linguistic grounds, and focus on explaining the remaining 4 frequencies through the corresponding equations.  This leads to 4 equations with 5 parameters (since $U_{S\to V}$ no longer enters the picture).  To balance the number of equations and parameters, we may fix the ratio of the parameters $U_{O\to S}$ and $U_{V\to S}$ (the ones that were not subject to degeneracy in [Figure \ref{fig:fig2}]), and solve by gradient descent.  The results are shown below [Figure \ref{fig:fig3}].  We notice that our predictions that $U_{V\to O}\gtrapprox U_{O\to V} > U_{S\to O}$ and $U_{V\to S} > U_{O\to S}$ are stable, while our previous prediction that $U_{S\to V}$ is very large vanishes because this unlock operation only plays a role in the OSV order, whose frequency we ignore. The ``$U_{S\to V}$ is very large" prediction was proposed as the cause of the small frequencies of OSV and OVS, a phenomenon which now has another causal explanation based on linguistic principles. It may still be that $U_{S\to V}$ would be large, presumably because this brain connection is rarely used, but a different model or experimental evidence may need to be employed to answer this question.

% \newpage

% \nocite{*}
\bibliographystyle{unsrt}
\bibliography{bibliography_svo.bib}

\begin{thebibliography}{10}

\bibitem{Dryer2005}
Matthew Dryer.
\newblock Order of subject, object, and verb.
\newblock In Haspelmath Martin, Dryer Matthew~S., Gil David, and Comrie
  Bernard, editors, {\em The World Atlas of Language Structures}, chapter~81,
  pages 330--333. Oxford University Press, Oxford, 2005.

\bibitem{Maurits2010}
Luke Maurits, Amy Perfors, and Daniel Navarro.
\newblock {Why are some word orders more common than others? A Uniform
  Information Density account}.
\newblock Technical report, 2010.

\bibitem{Maurits2011}
Luke Maurits.
\newblock {Representation , information theory and basic word order}.
\newblock Technical Report September, 2011.

\bibitem{Maurits2014}
Luke Maurits and Thomas~L. Griffiths.
\newblock {Tracing the roots of syntax with Bayesian phylogenetics}.
\newblock {\em Proceedings of the National Academy of Sciences of the United
  States of America}, 111(37):13576--13581, sep 2014.

\bibitem{Gell-Mann2011}
Murray Gell-Mann and Merritt Ruhlen.
\newblock {The origin and evolution of word order}.
\newblock {\em Proceedings of the National Academy of Sciences of the United
  States of America}, 108(42):17290--17295, oct 2011.

\bibitem{Krupa1982}
Viktor Krupa.
\newblock {Syntactic Typology and Linearization}.
\newblock Technical Report~3, 1982.

\bibitem{Hammarstrom2015}
Harald Hammarstr{\"{o}}m.
\newblock {The Basic Word Order Typology: An Exhaustive Study}.
\newblock Technical report, 2015.

\bibitem{Huber2013}
Magnus Huber and the~APiCS Consortium.
\newblock {Order of subject, object, and verb}.
\newblock {\em Atlas of Pidgin and Creole Language Structures Online}, pages
  330--333, 2013.

\bibitem{Aitchison1986}
Jean Aitchison.
\newblock {Word order universals. John A. Hawkins, (Quantitative Analysis of
  Linguistic Structure Series.) Academic Press, New York, London, Sydney,
  1983}, nov 1986.

\bibitem{Tomlin1988}
Russell~S. Tomlin.
\newblock {\em {Basic word order. Functional principles.}}
\newblock Number~1. Cambridge University Press (CUP), mar 1988.

\bibitem{Greenberg1963}
Joseph Greenberg.
\newblock {Some universals of grammar with particular reference to the order of
  meaningful elements}.
\newblock Technical report, 1963.

\bibitem{Friederici2017}
A.D. Friederici and N.~Chomsky.
\newblock {\em Language in Our Brain: The Origins of a Uniquely Human
  Capacity}.
\newblock The MIT Press. MIT Press, 2017.

\bibitem{Papadimitriou2020}
Christos~H. Papadimitriou, Santosh~S. Vempala, Daniel Mitropolsky, Michael
  Collins, and Wolfgang Maass.
\newblock Brain computation by assemblies of neurons.
\newblock {\em Proceedings of the National Academy of Sciences},
  117(25):14464--14472, 2020.

\bibitem{Dunn2011}
Michael Dunn, Simon~J. Greenhill, Stephen~C. Levinson, and Russell~D. Gray.
\newblock {Evolved structure of language shows lineage-specific trends in
  word-order universals}.
\newblock {\em Nature}, 473(7345):79--82, may 2011.

\bibitem{frankland2020}
Steven~M Frankland and Joshua~D Greene.
\newblock {Two Ways to Build a Thought: Distinct Forms of Compositional
  Semantic Representation across Brain Regions}.
\newblock {\em Cerebral Cortex}, 30(6):3838--3855, 04 2020.

\bibitem{Buzsaki2007}
György Buzsáki, Kai Kaila, and Marcus Raichle.
\newblock Inhibition and brain work.
\newblock {\em Neuron}, 56(5):771--783, 2007.

\bibitem{Jackson2015}
Georgina~M. Jackson, Amelia Draper, Katherine Dyke, Sophia~E. Pépés, and
  Stephen~R. Jackson.
\newblock Inhibition, disinhibition, and the control of action in tourette
  syndrome.
\newblock {\em Trends in Cognitive Sciences}, 19(11):655--665, 2015.

\bibitem{Letzkus2015}
Johannes J. Letzkus, Steffen B.E. Wolff, and Andreas Lüthi.
\newblock Disinhibition, a circuit mechanism for associative learning and
  memory.
\newblock {\em Neuron}, 88(2):264--276, 2015.

\bibitem{Laughlin1998}
Simon~B. Laughlin, Rob~R. de~Ruyter~van Steveninck, and John~C. Anderson.
\newblock The metabolic cost of neural information.
\newblock {\em Nature Neuroscience}, 1(1):36--41, May 1998.

\bibitem{Laughlin2001}
Simon~B Laughlin.
\newblock Energy as a constraint on the coding and processing of sensory
  information.
\newblock {\em Current Opinion in Neurobiology}, 11(4):475--480, 2001.

\bibitem{Attwell2001}
David Attwell and Simon~B. Laughlin.
\newblock An energy budget for signaling in the grey matter of the brain.
\newblock {\em Journal of Cerebral Blood Flow \& Metabolism},
  21(10):1133--1145, 2001.
\newblock PMID: 11598490.

\bibitem{Niven2008}
Jeremy~E. Niven and Simon~B. Laughlin.
\newblock {Energy limitation as a selective pressure on the evolution of
  sensory systems}.
\newblock {\em Journal of Experimental Biology}, 211(11):1792--1804, 06 2008.

\bibitem{sethna2006}
Sethna James.
\newblock {\em Statistical Mechanics : Entropy, Order Parameters and
  Complexity.}
\newblock Number Vol. 14 in Oxford Master Series in Physics. OUP Oxford, 2006.

\bibitem{Papadimitriou1993}
Christos~H. Papadimitriou and Mihalis Yannakakis.
\newblock The traveling salesman problem with distances one and two.
\newblock {\em Mathematics of Operations Research}, 18(1):1--11, 1993.

\end{thebibliography}

\end{document}